\documentclass[11pt]{article}
\PassOptionsToPackage{table}{xcolor}
\usepackage{acl}

\usepackage{graphicx}
\usepackage{pgfplots}
\pgfplotsset{compat=1.18}
\usepackage{times}
\usepackage{enumitem}
\usepackage{multirow}
\usepackage{url}
\usepackage{latexsym}
\usepackage{booktabs}
\usepackage{amsmath, amssymb}

\usepackage{bigstrut}
\usepackage{bbm}
\usepackage{bbding}
\usepackage{bbold}
\usepackage{mathtools}
\usepackage{tcolorbox}
\usepackage{mathptmx}
\usepackage{hyperref}
\usepackage{amsthm}
\usepackage{subcaption}
\newtheorem{definition}{Definition}

\usepackage[T1]{fontenc}

\ifPDFTeX
  \usepackage[utf8]{inputenc}
\fi
\usepackage{microtype}
\usepackage{listings}
\definecolor{mygreen}{rgb}{0,0.6,0}
\definecolor{mygray}{rgb}{0.5,0.5,0.5}

\newcommand{\incode}[1]{\lstinline{#1}}
\usepackage{inconsolata}
\usepackage{xspace} 
\usepackage[ruled,linesnumbered]{algorithm2e}
\SetAlgoNlRelativeSize{-1}
\usepackage{tablefootnote}
\newcommand{\sys}{{\upshape\textsc{Agent+P}}\xspace}

\usetikzlibrary{shapes.geometric, arrows.meta, positioning, calc, fit, backgrounds, shadows}
\definecolor{moduleBlue}{RGB}{218, 232, 252}
\definecolor{moduleBorder}{RGB}{108, 142, 191}
\definecolor{dataGreen}{RGB}{213, 232, 212}
\definecolor{dataBorder}{RGB}{130, 179, 102}
\definecolor{llmOrange}{RGB}{255, 230, 204}
\definecolor{llmBorder}{RGB}{215, 155, 0}

\title{Agent+P: Guiding UI Agents via
Symbolic Planning}

\author{
Shang Ma\textsuperscript{1},
Xusheng Xiao\textsuperscript{2},
Yanfang Ye\textsuperscript{1,$\dagger$} \\
\normalfont
\textsuperscript{1}University of Notre Dame \\
\normalfont
\textsuperscript{2}Arizona State University \\
\normalfont
\textsuperscript{$\dagger$}Corresponding Author \\
\normalfont
\texttt{\{sma5,yye7\}@nd.edu}, \texttt{xusheng.xiao@asu.edu}
}

\begin{document}
\maketitle
\begin{abstract}
Large Language Model (LLM)-based UI agents show great promise for UI automation but often hallucinate in long-horizon tasks due to their lack of understanding of the global UI transition structure. To address this, we introduce \sys, a novel framework that leverages symbolic planning to guide LLM-based UI agents. Specifically, we model an app's UI transition structure as a UI Transition Graph (UTG), which allows us to reformulate the UI automation task as a pathfinding problem on the UTG. This further enables an off-the-shelf symbolic planner to generate a provably correct and optimal high-level plan, preventing the agent from redundant exploration and guiding the agent to achieve the automation goals. \sys is designed as a plug-and-play framework to enhance existing UI agents. Evaluation on the AndroidWorld benchmark demonstrates that \sys improves the success rates of state-of-the-art UI agents by up to $14.32\%$ and reduces the action steps by $37.70\%$. Our code is available at:
\url{https://github.com/KiteFlyKid/agentp}.
\end{abstract}

\section{Introduction}

With mobile applications (apps) woven into all parts of our daily life, it is critically important to ensure the high quality of apps.
User interface (UI) automation, the process of programmatically executing sequences of UI interactions, has become an essential method for improving app quality by enabling automated testing for bug and vulnerability detection~\cite{lai2019goal,ma2024careful} and supporting user task automation~\cite{orru2023human, rawles2024androidworld,li2025mobileuse,zhang2023appagent}.  

While recent advances in UI automation, notably the integration of LLM-based UI agents~\cite{liu2025temac,zhao2024gui,ran2024guardian} that explicitly model the available actions in each UI screen, have demonstrated encouraging results, the ever-growing complexity of modern UIs continues to hinder effective and efficient automation.
In particular, existing approaches struggle in \textbf{long-horizon planning} tasks that require navigating via multiple UIs since such multi-step planning often leads to increased hallucination rates~\cite{liu2023llm+,wei2025plangenllms,xie2025mirage,wuatlas}.   
For example, the LLM-based agents employed by these approaches typically follow a depth-first strategy to find valid action sequences, making decisions based on local UI states without understanding the global transition structure, where different UIs can lead to distinct subsequent actions. Consequently, they often fail to derive valid sequences that accomplish task goals and repeatedly waste effort on actions that diverge from those goals.

\noindent\textbf{Key Insights.} 
To address this fundamental limitation,\textit{ we introduce an external planner module that provides LLM-based UI agents with global transition knowledge extracted through program analysis. This module leverages established planning algorithms to prevent redundant exploration and guide the agent toward diverse action sequences that are more likely to achieve the automation goals}.
Specifically, we model an app's global transition  structure via a UI Transition Graph (UTG)~\cite{sun2025gui,wen2024autodroid}, with nodes representing UIs and edges representing user-triggered UI transitions. 
A UI automation problem, navigating from a start UI to a target UI, can thus be formulated as a \textit{pathfinding problem} on the UTG, where the objective is to find an optimal path from the start node to the target node. 
This formalization enables the use of off-the-shelf symbolic planners to derive provably correct and optimal plans, thereby fundamentally mitigating high-level planning hallucinations and enhancing the reliability of LLM-based UI agents.

\noindent\textbf{Our Method.} 
Building upon these insights, in this paper, we introduce \sys, an agentic framework that leverages symbolic planning to guide UI agents. Given a natural language UI automation goal and the app for automation, \sys operates iteratively in four stages until the goal is achieved: the \textbf{UTG Builder} constructs a static UTG of the app and dynamically updates it during automation. The \textbf{Node Selector} maps the natural language goal to a targeted node in the UTG. 
The \textbf{Plan Generator} translates the UTG into a pathfinding problem using Planning Domain Definition Language (PDDL)~\cite{aeronautiques1998pddl}, which is then solved by an external symbolic planner. The resulting symbolic plan is subsequently converted into natural language instructions. Finally, the \textbf{UI Explorer} interacts with the app to execute the translated instructions to navigate to the goal.

\sys is designed as a plug-and-play planning framework that can be incorporated with and enhance existing UI agents. We evaluate \sys by integrating it with four state-of-the-art agents on the AndroidWorld benchmark. Our results demonstrate that by leveraging symbolic planning on the UTG, \sys increases the success rates of baseline agents by up to $14.32\%$ and reduces the action steps by $37.70\%$.

Our primary contributions are as follows:
\begin{itemize}[noitemsep, topsep=1pt, partopsep=1pt, leftmargin=*]

    \item We propose \sys, a novel framework that leverages symbolic planning to provide LLM-based UI agents with global transition information derived from program analysis, mitigating the long-horizon planning failures commonly faced by these agents.
    \item We present a novel formalization that maps the problem of UI automation into a pathfinding problem in the UTG, making it solvable with provably correct symbolic planners.
    \item We conduct extensive evaluation on the AndroidWorld benchmark, demonstrating that \sys substantially improves the success rate and efficiency of four state-of-the-art UI agents.
\end{itemize}

\section{Background and Motivation}

\subsection{UI and UI Transition Graph}
To motivate our method, we begin by introducing the concepts of widget, UI, and the modeling of UI  transitions (i.e., UTG). 

\begin{definition}[Widget, Action]
\label{def:widget}
\emph{A} widget, \emph{denoted as $w$, is a basic interactive element on a UI screen. An} action, \emph{denoted as $a$, is a 2-tuple $a=(w, e)$, where $w$ is a widget, $e$ is the user event (e.g., \texttt{click}, \texttt{input}).}
\end{definition}

Following existing UI agents that represent a UI as a sequence of widgets and supported actions~\cite{androidworldleaderboard,ye2025mobile,dai2025advancing,li2025mobileuse},  
we define the UI state (shortened as UI) as follows:
\begin{definition}[UI]
\label{def:ui}
\emph{A} UI, \emph{denoted as $u$, is an n-tuple of all unique actions available on the screen, $u=(a_1, a_2, \ldots, a_n)$, where $n$ is the total number of available actions.} 
\end{definition}
\noindent This level of abstraction is sufficient to represent UI transitions while avoiding state explosion~\cite{valmari1996state}.

\begin{definition}[UI Transition Graph]
\label{def:utg}
A UTG for an app is a directed graph $G = (\mathcal{U}, \mathcal{T}, \epsilon)$ that models the transition structure of the app. 
\end{definition}

\begin{itemize}[noitemsep, topsep=1pt, partopsep=1pt, leftmargin=*]
    \item $\mathcal{U}$ is a finite set of nodes, where each node $u \in \mathcal{U}$ represents a UI $u$ in the app.
    \item $\mathcal{T} \subseteq \mathcal{U} \times \mathcal{U}$ is a set of directed edges. An edge $(u_i, u_j) \in \mathcal{T}$ represents a transition from UI $u_i$ to UI $u_j$. 
    \item $\epsilon : \mathcal{T} \rightarrow \mathcal{A}$ is an edge-labeling function. It maps each transition $(u_i, u_j)$ to the action $a = (w, e)$ that triggers it, where the widget $w$ is an element of the source UI $u_i$.
\end{itemize}

\autoref{fig:utgCalendar} shows an example of UTG of an Android app named Simple Calendar Pro in AndroidWorld benchmark.

\begin{table}[t!]
\centering
\fontsize{10}{12}\selectfont
\setlength{\tabcolsep}{3pt}
\begin{tabular}{@{}lcccc@{}}
\toprule
& \multicolumn{2}{c}{\textbf{Nodes}} & \multicolumn{2}{c}{\textbf{Edges}} \\
\cmidrule(lr){2-3} \cmidrule(lr){4-5}
\textbf{Agent} & $>$ & $\le$ & $>$ & $\le$ \\ \midrule
DroidRun    & 27.0 & 71.2 & 62.7 & 168.0 \\
LX-GUIAgent & 29.3 & 53.8 & 65.9 & 129.2 \\
AutoGLM     & 39.4 & 42.0 & 90.6 & 100.3 \\
Finalrun    & 23.8 & 55.0 & 59.5 & 125.6 \\
UI-Venus    & 18.0 & 50.7 & 66.0 & 108.0 \\ \bottomrule
\end{tabular}
\caption{Average UTG node and edge counts. Headers compare app-specific success rate ($\text{SR}_{\text{app}}$) with the overall average ($\overline{\text{SR}}$).}
\label{tab:utg_comparison_separated}
\end{table}

\subsection{Motivational Study}
\label{subsec:motivationalstudy}

To investigate how UI complexity affects agent performance, we conduct a motivational study using the AndroidWorld benchmark~\cite{rawles2024androidworld}. This benchmark consists of 116 programmatic tasks across 20 Android apps, where an agent must navigate a given app to satisfy a natural language instruction.

Performance is evaluated using a task-level ``success rate''. While published results typically report a single average across all tasks, we aim to uncover performance variations across different applications. We define $\overline{\text{SR}}$ as the agent's overall average success rate across the entire benchmark. For any given app, we calculate $\text{SR}_{\text{app}}$, the success rate specific to that app.

Based on these metrics, we classify apps into two categories: those where the agent exceeds its average performance ($\text{SR}_{\text{app}} > \overline{\text{SR}}$) and those where it underperforms ($\text{SR}_{\text{app}} \le \overline{\text{SR}}$).

\autoref{tab:utg_comparison_separated} compares the UI complexity, measured by the number of UTG nodes and edges, between these two categories. Our statistical analysis shows that apps where agents underperform ($\text{SR}_{\text{app}} \le \overline{\text{SR}}$) have significantly more UTG nodes and edges than those where they succeed ($p=0.03$). This indicates that existing agents struggle with high-complexity UIs, motivating our approach to leverage the app's transition structure.

\section{Problem Formulation}
In this section, we first define the problem of targeted UI automation, and how to convert it into an equivalent classical planning problem.

\subsection{Problem Definition}
With the structure of the app modeled as a UTG, we can now formally define the task of UI automation.

\begin{definition}[UI Automation]
\label{def:gui_automation}
UI automation is the process of programmatically executing a sequence of actions $\pi = \langle a_1, a_2, \dots, a_N \rangle$, where each action $a_i \in \mathcal{A}$.
\end{definition}

In this work, we focus on a specific, goal-oriented variant of this task.

\begin{definition}[Targeted UI Automation]
\label{def:targeted_gui_automation}
 Given an app with an initial UI $u_{init}$ and a target UI $u_{target}$, the objective is to find a valid sequence of actions $\pi = \langle a_1, a_2, \dots, a_N \rangle$, where $a_i = (w_i, e_i)$, that navigates the app from $u_{init}$ to $u_{target}$.
\end{definition}

\textbf{We formulate targeted UI automation as a pathfinding problem on the UTG}. 
Let $\kappa: \mathcal{A} \rightarrow \mathbb{R}^+$ be a cost function that assigns a 
positive cost to each action, representing computational resources, execution time, 
or other relevant metrics. The problem is then to find a path from $u_{init}$ to $u_{target}$ 
that minimizes the total execution cost:
\[
\pi^* = \arg\min_{\pi} \sum_{i=1}^{N} \kappa(a_i).
\]
To simplify the formulation, in this work, we adopt a uniform action cost. This reduces the cost-minimization task to the classical shortest path problem, where the objective is to find the path from $u_{init}$ to $u_{target}$ with the fewest actions.

\begin{figure}[t]
\centering
\begin{lstlisting}[language=lisp, caption={Domain PDDL for targeted UI automation.}, label={lst:domain_pddl}]
(define (domain utg-automation)
  (:requirements :strips :typing)
  (:types node - object)
  (:predicates
    (at ?n - node)
    (connected ?from - node ?to - node)
    (visited ?n - node) (goal-node ?n - node)
    (goal-achieved ?n - node))
  (:action navigate
    :parameters (?from - node ?to - node)
    :precondition (and
      (at ?from) (connected ?from ?to))
    :effect (and
      (not (at ?from)) (at ?to)
      (visited ?to)
      (when (goal-node ?to)
        (goal-achieved ?to)))))
\end{lstlisting}
\end{figure}
\begin{table*}[t]
\centering
\resizebox{\textwidth}{!}{
\begin{tabular}{@{}ll@{}}
\toprule
\textbf{UI Automation Notation}                                               & \textbf{Classical Planning Equivalent}                               \\ \midrule
A UI state (UI) $u \in \mathcal{U}$                                           & A state $s \in \mathcal{S}$ where the predicate $at(u)$ is true     \\
The set of all UIs $\mathcal{U}$                                              & The state space $\mathcal{S}$                                       \\
The initial UI  $u_{init}$                                                        & The initial state $s_{init}$, defined by $at(u_{init})$             \\
The target UI $u_{target}$                                                        & The goal $\mathcal{G}$, specified by the condition $at(u_{target})$ \\
A UI transition via the edge $(u_i, u_j)$ with label $a$                & A planning action                                                   \\
\multicolumn{1}{r}{\textit{Precondition: the app is on UI $u_i$}}             & $pre(a): at(u_i)$                                                    \\
\multicolumn{1}{r}{\textit{Effect: the app moves to UI $u_j$}}            & $eff(a): \lnot at(u_i) \land at(u_j)$                                \\
A path from $u_{init}$ to $u_{target}$ via the sequence of action $\pi$ & A plan $\pi$                                                        \\ \bottomrule
\end{tabular}
}
\caption{Mapping of UI automation notation to classical planning equivalent.}
\label{tab:utg_planning_mapping}
\end{table*}

\subsection{Symbolic and Classical Planning}
Symbolic planning is a long-standing area of AI concerned with finding a sequence of actions to achieve a predefined goal. The most fundamental and widely studied form is \textbf{classical planning} where a planning problem instance, $P$, is formally described as a tuple $P = \langle \mathcal{D}, s_{init}, \mathcal{G} \rangle$, where $\mathcal{D} = \langle \mathcal{F}, \mathcal{A} \rangle$ is the planning domain. These components are defined as follows:
\begin{itemize}[noitemsep, topsep=1pt, partopsep=1pt, leftmargin=*]
    \item \textbf{States}: $\mathcal{F}$ is a set of fluents or predicates that describe the properties of the world. A state $s$ is a complete assignment of truth values to all fluents in $\mathcal{F}$. The set of all possible states is the state space $\mathcal{S}$.
    \item \textbf{Initial State}: $s_{init} \in \mathcal{S}$ is the initial state of the world.
    \item \textbf{Goal}$: \mathcal{G}$ is the goal specification, a set of conditions on states. Any state $s \in \mathcal{S}$ that satisfies all conditions in $\mathcal{G}$ is a goal state.
    \item \textbf{Actions}: $\mathcal{A}$ is a set of actions. Each action $a \in \mathcal{A}$ is defined by its preconditions, $\text{pre}(a)$, and its effects, $\text{eff}(a)$. An action can only be executed in a state where its preconditions are met, and its effects describe how the state changes after its execution.
\end{itemize}
A  plan, $\pi$, is a sequence of actions $\langle a_1, a_2, \dots, a_N \rangle$ that transforms the initial state $s_{init}$ into a goal state. This is achieved by applying the actions sequentially, where each action $a_i$ is applicable in the state resulting from the execution of $a_{i-1}$, and the final state after executing $a_N$ satisfies $\mathcal{G}$.

The Planning Domain Definition Language (PDDL)~\cite{aeronautiques1998pddl} is the standard language for representing such planning problems, typically using two files: a domain file defining $\mathcal{F}$ and $\mathcal{A}$, and a problem file defining $s_{init}$ and $\mathcal{G}$.

\begin{figure*}[t!]
    \centering
    \includegraphics[width=0.95\textwidth]{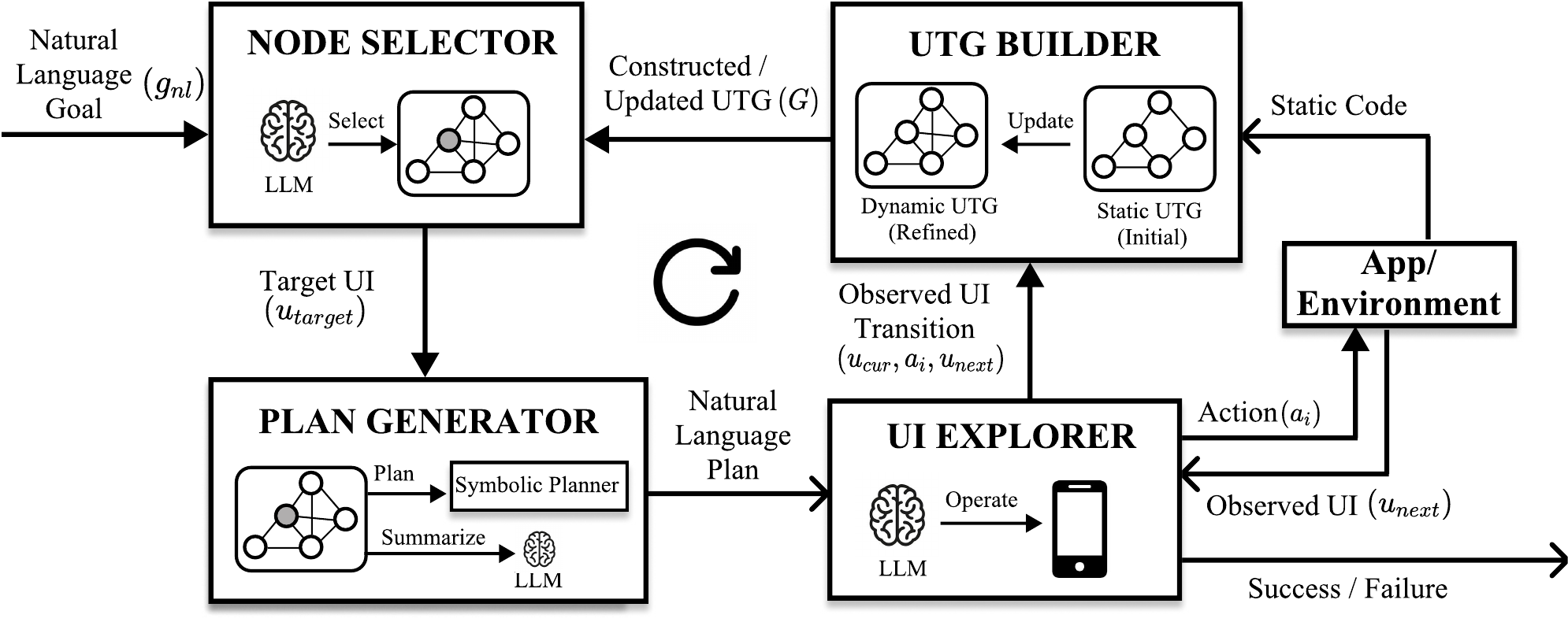}
    \caption{Overview of \sys.}
    \label{fig:overview}
\end{figure*}

\subsection{UI Automation  to Classical Planning}
The pathfinding problem can be naturally cast into a classical planning problem, allowing us to leverage classical planners to compute the solution, i.e., the shortest path.  \autoref{tab:utg_planning_mapping} illustrates the mapping from the UI automation domain to the classical planning domain, which is elaborated as follows:
\begin{itemize}[noitemsep, topsep=1pt, partopsep=1pt, leftmargin=*]
    \item \textbf{States}: A planning state $s$ corresponds to the app being at a specific UI $u$. We can define a predicate $at(u)$ which is true if the app is currently on UI $u \in \mathcal{U}$. The state space $\mathcal{S}$ is the set of all possible UIs, $\mathcal{U}$.
    \item \textbf{Initial State}: The initial state $s_{init}$ is defined by the predicate $at(u_{init})$ being true.
    \item \textbf{Goal}: The goal $\mathcal{G}$ is specified by the condition that the predicate $at(u_{target})$ must be true.
    \item \textbf{Actions}: For each UI transition $(u_i, u_j)$ in the UTG triggered by 
    an action $a = (w, e)$, we define a planning operator with precondition $at(u_i)$ 
    and effects $\lnot at(u_i)$ and $at(u_j)$.
\end{itemize}
Following this formulation, we define a general PDDL domain file template applicable to any targeted UI automation task, as shown in \autoref{lst:domain_pddl}. Any specific user task, represented by UTG with a specified target UI, $u_{target}$, can be translated into a corresponding PDDL problem file via this template. For instance, to solve the ``Change the time zone'' task in the Simple Calendar Pro app, the UTG from \autoref{fig:utgCalendar} is converted into the problem file presented in \autoref{lst:problem_pddl}. This symbolic representation allows us to employ a classical planner to efficiently compute a valid and optimal sequence of actions to navigate the app from the initial UI, $u_{init}$, to the target, $u_{target}$.

\section{\sys}
\label{sec:methodology}

\let\oldnl\nl%
\newcommand{\nonl}{\renewcommand{\nl}{\let\nl\oldnl}}%

\autoref{fig:overview} illustrates the overall architecture of \sys, while the step-by-step workflow is detailed in Algorithm~\ref{alg:sys}. 
Specifically, given a natural language goal specified by the user, \sys operates through four primary modules that interact in a continuous loop until the goal is achieved (success or failure): the \textbf{UTG Builder}, the \textbf{Node Selector}, the \textbf{Plan Generator}, and the \textbf{UI Explorer}.  
In the following subsections, we elaborate on the design rationale and functionality of each module.

\subsection{UTG Builder}
 Existing methods for constructing a UTG rely on either dynamic analysis~\cite{wen2024autodroid,sun2025gui}, which is accurate but often suffers from high cost and incomplete coverage, or static analysis~\cite{dfsandroid,yang2018static}, which is more comprehensive but can introduce infeasible transitions~\cite{liu2022promal}.

To overcome these limitations, \sys utilizes a hybrid approach that synergizes both techniques to build UTG. \sys begins by performing static analysis to construct an initial UTG (Line 1 of Algorithm~\ref{alg:sys}), following established practices that track API calls responsible for UI transitions in the app's source code~\cite{yang2018static,liu2022promal}.  

Based on the environment feedback, the initial UTG is dynamically verified and refined as the UI Explorer explores the app (Line 11 and Line 17)~\cite{zhu2025language}.
Specifically, let the current UTG be $G = (\mathcal{U}, \mathcal{T}, \epsilon)$. During UI automation, an observed transition $(u_i, a_{obs}, u_j)$, where $u_i \in \mathcal{U}$, updates the graph to $G' = (\mathcal{U}', \mathcal{T}', \epsilon')$ in one of three ways:

\begin{itemize}[noitemsep, topsep=1pt, partopsep=1pt, leftmargin=*]
    \item \textbf{Update Edge:} An action leads from an existing source node to a target node, but the recorded action does not match the corresponding one in the UTG. Formally, if $(u_i, u_j) \in \mathcal{T}$ and $\epsilon((u_i, u_j)) \neq a_{obs}$, the labeling function is updated such that $\epsilon'((u_i, u_j)) = a_{obs}$, while $\mathcal{U}' = \mathcal{U}$ and $\mathcal{T}' = \mathcal{T}$.

    \item \textbf{Add Edge:} An action creates an edge between two nodes in the UTG. This occurs when $u_j \in \mathcal{U}$ and $(u_i, u_j) \notin \mathcal{T}$. A new transition is added by setting $\mathcal{T}' = \mathcal{T} \cup \{(u_i, u_j)\}$ and extending $\epsilon'$ with $\epsilon'((u_i, u_j)) = a_{obs}$.

    \item \textbf{Add Node:} An action leads to a UI that is not yet in the UTG. If $u_j \notin \mathcal{U}$, a new node and edge are added to the graph: $\mathcal{U}' = \mathcal{U} \cup \{u_j\}$, $\mathcal{T}' = \mathcal{T} \cup \{(u_i, u_j)\}$, and $\epsilon'$ is extended with $\epsilon'((u_i, u_j)) = a_{obs}$.
\end{itemize}

This approach allows \sys to maintain a UTG that is both comprehensive and dynamically accurate, combining the breadth of static analysis with the precision of real-time exploration.

\begin{algorithm}[t!]
\caption{Workflow of \sys}
\label{alg:sys}
\small
\SetAlgoLined
\SetAlgoNoEnd

\KwIn{Natural language goal $g_{nl}$, App $\mathcal{A}$, Max running steps $maxStep$}
\KwOut{Automation outcome: \textit{Success} or \textit{Failure}}
\nonl \textit{Aliases: \textbf{UB} $\leftarrow$ UtgBuilder; \textbf{NS} $\leftarrow$ NodeSelector; \textbf{PG} $\leftarrow$ PlanGenerator; \textbf{UE} $\leftarrow$ UiExplorer}

$G\gets \text{UB.buildStaticUTG}(\mathcal{A})$\;
$u_{cur}\gets \text{getCurrentUI}(\mathcal{A})$\;
$steps\gets0$\;
$u_{target}\gets \varnothing$\;
\While{ $steps\leq maxStep$}{
    $u_{target}\gets \text{NS.selectTargetNode}(g_{nl}, G,u_{target})$\;
    $Plan\gets \text{PG.generatePlan}(u_{cur}, u_{target}, G)$\;
    \uIf{$Plan$ is valid}{
        \For{each action $a_i$ in $Plan$}{
            $u_{next}\gets \text{UE.act}(a_i)$\;
            $G\gets \text{UB.update}(G, u_{cur}, a_i, u_{next})$\;
            $u_{cur}\gets u_{next}$\;
        }
    }
    \Else{
        $neighbors\gets \text{PG.getNeighbors}(u_{cur}, G)$\;
        $a\gets \text{UE.decideAction}(neighbors, u_{cur})$\;
        $u_{next}\gets \text{UE.act}(a)$\;
        $G\gets \text{UB.update}(G, (u_{cur}, a, u_{next}))$\;
        $u_{cur}\gets u_{next}$\;
    }
    \If{$\text{UE.evaluate}(g_{nl})$}{
    \Return{\textit{Success}}\;
    }
    $steps \gets steps+1$\;
}
\Return{\textit{Failure}}\;
\end{algorithm}

\subsection{Node Selector}

The goal of the Node Selector is to map an unstructured, natural language goal $g_{nl}$ (e.g., ``Create a playlist'') to a concrete UI state ($u_{target}$). To bridge the semantic gap between user intent and UI elements~\cite{baechler2024screenai,li2021screen2vec,li2025ui}, we implement a hierarchical identification strategy based on both the reasoning capabilities of LLM and efficient semantic retrieval.

\noindent\textbf{MLLM-based Identification}. We query a Multimodal Large Language Model (MLLM) with the user's goal and the representations of candidate UIs. The MLLM is prompted to analyze the screen content and directly identify which UI screen corresponds to the user's objective. The specific prompt template used for this reasoning process is detailed in \autoref{lst:prompt_nodeSelector}.

\noindent\textbf{Embedding-based Fallback}. In cases where the MLLM fails to return a confident result, \sys computes semantic embeddings for both the user's input query and the textual representations of all UI nodes in the UTG. We then calculate the cosine similarity between the query embedding and each node embedding, selecting the node with the highest similarity score as the target.

\noindent\textbf{Dynamic Reflection.}
To reduce target-node selection errors, \sys does not rely on a single one-shot prediction. Instead, the Node Selector is invoked inside the execution loop (Line 6 in Algorithm~\ref{alg:sys}), allowing \sys to revise the target node at each planning iteration based on the updated UTG.

\subsection{Plan Generator}
Once the target node $u_{target}$ is identified, the task becomes a classical planning problem: finding the shortest sequence of actions from the current UI state to the target state. The Plan Generator orchestrates this by first converting the UTG into the PDDL format. It then invokes a classical planner to solve for an optimal PDDL plan (Line 7).

This symbolic plan is subsequently translated back to a sequence of clear, natural language instructions for the UI Explorer to execute. An example of a raw PDDL plan and its corresponding natural language translation are shown in Appendix in \autoref{lst:plan_pddl} and \autoref{tab:utgGuidePrompt}, respectively. In cases where the classical planner fails to find a valid path (e.g., if the target is unreachable), \sys implements a fallback strategy. Instead of a plan, it generates a textual summary of the k-hop neighbors from the current UI, providing contextual information to help the agent decide on its next steps (Line 14-16).

\subsection{UI Explorer}
The UI Explorer acts as the \sys's execution engine, emulating user interactions with the app (Line 10, Line 15-16, and Line 19-20).  It takes the natural language plan, and executes each step programmatically. After each action, it checks whether the resulting UI state matches the expected goal. 

A key design feature of the UI Explorer is its modularity. It is a plug-and-play component, allowing \sys to be integrated with various types of UI agents, including those based on LLMs, MLLMs, or other specialized models that incorporate capabilities like reflection and grounding. This flexibility is demonstrated in our evaluation (Section~\ref{sec:evalresults}), where we integrate \sys with different UI agents to showcase its  applicability.

\section{Evaluation Setting}
\label{sec:evalsetting}

\subsection{Datasets}
 We evaluate \sys in two distinct UI automation scenarios:  user task execution and automated UI testing.

\noindent\textbf{User Task Execution.} We use AndroidWorld~\cite{rawles2024androidworld}, which is a standard benchmark for evaluating how well UI agents perform real-world tasks. It provides automatic metrics to assess the success rate of an agent.
For this scenario, we evaluate a subset of apps from the AndroidWorld benchmark. This subset consists of apps where baseline agents most frequently failed, as identified in our motivational study in \autoref{subsec:motivationalstudy}. \autoref{tab:evaldataset} summarizes the characteristics of these apps, including the number of tasks and the complexity of their UTGs (nodes and edges).
The results on the full benchmark is provided in ~\autoref{tab:main_results_fullbenchmark}.

\noindent\textbf{Automated UI Testing}. No established benchmark currently exists. Therefore, we design a custom test case that requires the agent to navigate to an app's privacy policy page.  We choose this case for two primary reasons: (1) \textit{Relevance}: Every app is mandated to include a privacy policy page, yet these pages often require security auditing for  violations~\cite{yang2022describectx,slavin2016toward}, creating a strong need for automated UI testing. (2) \textit{Complexity}: The visual implementation and location of entry points (e.g., settings icons, hamburger menus) vary significantly across apps, posing a non-trivial navigation challenge for agents attempting to locate the target UI without errors.
For this scenario, we evaluate on all the apps from the AndroidWorld benchmark.

\subsection{Baselines} 
\label{subsec:baselines}
We compare \sys against two categories of state-of-the-art approaches in UI automation:

\noindent\textbf{General UI Agents.} We select \textit{DroidRun}, \textit{MobileUse}, and \textit{T3A}, which are agents from the official AndroidWorld leaderboard~\cite{androidworldleaderboard}. We chose these specific agents because other leaderboard entrants are either not open-source or their released code is non-functional. For our evaluation, we integrate each as the UI Explorer module within \sys. Note that T3A is tightly coupled with the AndroidWorld infrastructure and cannot be adapted for the UI testing scenario.
We also integrate \sys with the frontier computer use agent (CUA) from OpenAI~\cite{openaicomputeruse}.  

\noindent\textbf{UI Testing Tools.} We also compare against three approaches specifically for automated UI testing:

    \begin{itemize}[noitemsep, topsep=1pt, partopsep=1pt, leftmargin=*]

        \item \textit{Guardian~\cite{ran2024guardian}:} An LLM-based UI agent for automated testing with traditional planning.
        \item \textit{AutoDroid~\cite{wen2024autodroid}:} An LLM-based UI agent for UI Testing that utilizes UTGs without symbolic planning.
        \item \textit{GoalExplorer~\cite{lai2019goal}:} A fully symbolic UI automation approach based on UTGs.
    \end{itemize}
    Since these three approaches do not support the AndroidWorld environment, we evaluate them exclusively on the automated UI testing scenario.

\subsection{Metrics}
Consistent with prior work~\cite{rawles2024androidworld,li2025mobileuse}, we employ \textit{Success Rate} (SR) as the primary metric to evaluate the effectiveness of each UI automation approach. 

Since the integration of a symbolic planner introduces planning latency, it is critical to assess whether the improved effectiveness justifies this cost.  Therefore, to evaluate the overall efficiency, we report the average number of \textit{Steps} (UI actions) and the total \textit{Time} required to complete a task.

\begin{table}[t]
\centering
\fontsize{10}{12}\selectfont
\setlength{\tabcolsep}{4pt}
\begin{tabular}{@{}lrrr@{}}
\toprule
\multicolumn{4}{c}{\textbf{User Task Execution}} \\
\midrule
\textbf{Approach} & \textbf{SR (\%)} & \textbf{Steps} & \textbf{Time (s)} \\
\midrule
DroidRun &  15.91 & 5.18 & 33.38 \\
\rowcolor{gray!15}
\hspace{1em} + \sys & \textbf{28.41} & 17.09 & 99.50 \\
MobileUse & 9.09 & 20.45 & 256.02 \\
\rowcolor{gray!15}
\hspace{1em} + \sys & \textbf{11.36} & \textbf{19.61} & \textbf{219.28} \\
T3A & 24.32 & 17.89 & 114.39 \\
\rowcolor{gray!15}
\hspace{1em} + \sys & \textbf{38.64} & \textbf{13.43} & \textbf{95.90} \\
OpenAI CUA &  58.14 & 22.43 & 325.12 \\
\rowcolor{gray!15}
\hspace{1em} + \sys & \textbf{62.79} & \textbf{12.63} & 330.01 \\
\midrule
\multicolumn{4}{c}{\textbf{Automated UI Testing}} \\
\midrule
\textbf{Approach} & \textbf{SR (\%)} & \textbf{Steps} & \textbf{Time (s)} \\
\midrule
DroidRun & 60.00 & 6.10 & 19.90 \\
\rowcolor{gray!15}
\hspace{1em} + \sys & \textbf{60.00} & \textbf{3.80} & \textbf{16.90} \\
MobileUse & 40.00 & 7.36 & 22.32 \\
\rowcolor{gray!15}
\hspace{1em} + \sys & \textbf{50.00} & \textbf{4.00} & \textbf{18.45} \\
OpenAI CUA &  60.00 & 6.10 & 30.45 \\
\rowcolor{gray!15}
\hspace{1em} + \sys & \textbf{100.00} & \textbf{5.40} & \textbf{28.62} \\
AutoDroid & 35.00 & 9.42 & 46.43  \\
Guardian & 35.00 & 12.31 & 128.37  \\
GoalExplorer & 15.00 & 3.60 & 8.40  \\
\bottomrule
\end{tabular}
\caption{\sys compared with existing UI agents and traditional UI Automation approaches. We highlight the rows where \sys is integrated. SR is short for Success Rate.}
\label{tab:main_results}
\end{table}

\section{Evaluation Results}
\label{sec:evalresults}

\autoref{tab:main_results} presents our evaluation results across two UI automation scenarios. 

\subsection{User Task Execution}
\label{subsec:usertasks}

\noindent\textbf{Effectiveness.} As detailed in the upper section of~\autoref{tab:main_results}, \sys improves the success rate for all baseline agents. The integration yields the most substantial gains for \textit{DroidRun}, where the success rate nearly doubles from $15.91\%$ to $28.41\%$. Similarly, \textit{T3A} achieves a $14.32\%$ absolute improvement ($24.32\% \rightarrow 38.64\%$). \textit{OpenAI CUA} already performs strongly as a standalone agent ($58.14\%$), and \sys further improves it to $62.79\%$, achieving the best overall result. These results confirm that augmenting UI agents with global transition information effectively resolves navigation bottlenecks.

\noindent\textbf{Efficiency.} The impact of \sys varies depending on the agent's baseline behavior. For \textit{T3A}, \textit{MobileUse}, and \textit{OpenAI CUA}, \sys reduces the average steps, indicating that the planner helps agents avoid redundant exploration and solve tasks more directly. Conversely, for \textit{DroidRun}, we observe an increase in average steps ($5.18 \rightarrow 17.09$) and total time. This is because standalone \textit{DroidRun} tends to fail quickly on complex tasks, whereas \sys enables it to navigate deeper and complete harder tasks.

\subsection{Automated UI Testing}
\label{subsec:uitesting}

\noindent\textbf{Effectiveness.} As shown in the lower section of~\autoref{tab:main_results}, general UI agents surpass dedicated UI testing tools. Specifically, \textit{DroidRun + \sys} achieves a success rate of $60.00\%$, outperforming \textit{AutoDroid} and \textit{Guardian} by $25.00\%$. \textit{OpenAI CUA + \sys} achieves the best result with a $100.00\%$ success rate, improving over standalone \textit{OpenAI CUA} by $40.00\%$. In contrast, the purely symbolic approach \textit{GoalExplorer} struggles ($15.00\%$).

\noindent\textbf{Efficiency.} While the integration of \sys yields varying improvements in effectiveness (boosting \textit{MobileUse} by roughly $10\%$ while \textit{DroidRun} remains consistent), it drastically reduces the effort required to reach the target UI.  Since the privacy policy page is a fixed target, the step count provides a direct measure of navigation optimality.  \textit{DroidRun + \sys} reduces the average path length from $6.10$ steps to $3.80$ steps (a $37.70\%$ reduction) and total time from $19.90$s to $16.90$s compared to the standalone configuration. Similarly, \textit{MobileUse + \sys} reduces the average steps from $7.36$ to $4.00$. \textit{OpenAI CUA + \sys} also reduces the steps from $6.10$ to $5.40$ while reaching $100.00\%$ success. These results demonstrate that while standalone agents often rely on exploration or trial-and-error to find a specific page, \sys guides them via the optimal path derived from the underlying UTG, minimizing redundant interactions and accelerating the testing process.

\subsection{Ablation Study}
\label{subsec:ablation}

\begin{table}[t]
\centering
\small
\setlength{\tabcolsep}{4pt}
\renewcommand{\arraystretch}{0.95}
\begin{tabular}{@{}llr@{}}
\toprule
\textbf{Component} & \textbf{Variant} & \textbf{SR (\%)} \\
\midrule
UTG Builder
    & w/o Dynamic Update & 34.09 \\
\multirow{2}{*}{Node Selector}
    & w/o Embedding & 36.36 \\
    & w/o MLLM & 27.27 \\
\multirow{2}{*}{Plan Generator}
    & DFS-based & 22.73 \\
    & BFS-based & 25.00 \\
\midrule
\multicolumn{2}{l}{\textbf{\sys (Full Method)}}
    & \textbf{38.64} \\
\bottomrule
\end{tabular}
\caption{Ablation results for \sys.}
\label{tab:ablation_results}
\end{table}

To validate the design choices within \sys, we evaluate the individual contribution of its three core components: the UTG Builder, Node Selector, and Plan Generator. \autoref{tab:ablation_results} summarizes the success rates of \sys (integrated with T3A) compared to variants where specific features are removed or replaced.

\noindent\textbf{Impact of UTG Builder.}
We further examine the UTG Builder by disabling the dynamic update mechanism (``w/o Dynamic Update''). This variant, which relies on a static or append-only graph without refining existing transitions, achieves an SR of 34.09\%. The drop in performance confirms that dynamically updating the UTG is beneficial for maintaining an accurate UTG.

\noindent\textbf{Impact of Node Selector.}
We investigate the role of the Node Selector by ablating its hierarchical strategies.
Removing the MLLM (``w/o MLLM'') results in a sharp decline in performance to $27.27\%$. This underscores that the visual and semantic reasoning capabilities of the MLLM are critical for correctly identifying relevant UI elements.
Removing the embedding-based feedback (``w/o Embedding'') yields a smaller reduction ($36.36\%$), indicating that while embeddings help refine the selection, the MLLM provides the primary guidance.

\noindent\textbf{Impact of Plan Generator.}
We first assess the effectiveness of our symbolic planning approach by replacing it with standard graph search algorithms: Depth-First Search (DFS) and Breadth-First Search (BFS), which traverse the UTG in either DFS or BFS order, respectively, and generate guidance paths containing the target node. As shown in ~\autoref{tab:ablation_results}, replacing the symbolic planner leads to a huge performance drop, with DFS and BFS achieving only $22.73\%$ and $25.00\%$ SR, respectively. This degradation occurs because standard search algorithms blindly explore the large state space of Android apps, often exhausting the time budget or step limit before achieving the goal. In contrast, our symbolic planner leverages the high-level logic of the UTG to prune irrelevant paths, achieving a substantially higher SR of $38.64\%$.

\begin{figure}[t]
\centering
\begin{tikzpicture}
\begin{axis}[
    width=0.82\columnwidth,
    height=3.0cm,
    xmin=0.8, xmax=4.2,
    ymin=89.5, ymax=101,
    xtick={1,2,3,4},
    ytick={90,95,100},
    yticklabel={\pgfmathprintnumber{\tick}\%},
    xlabel={Step},
    ylabel={Accuracy},
    label style={font=\small},
    tick label style={font=\small},
    axis lines=left,
    grid=major,
    grid style={dashed,gray!30},
    clip=false
]
\addplot[thick,black!65,mark=*,mark size=2pt]
coordinates {
(1,90.91)
(2,95.45)
(3,97.73)
(4,100.00)
};

\node[font=\scriptsize,above] at (axis cs:1,90.70) {90.91\%};
\node[font=\scriptsize,above] at (axis cs:2,95.35) {95.45\%};
\node[font=\scriptsize,above] at (axis cs:3,97.67) {97.73\%};
\node[font=\scriptsize,above left] at (axis cs:4,100) {100\%};
\end{axis}
\end{tikzpicture}
\caption{Node Selector accuracy.}
\label{fig:accuracynodeselector}
\end{figure}

\noindent\textbf{Accuracy of Node Selector.}
As shown in
\autoref{fig:accuracynodeselector}, the Node Selector correctly identifies the target for 40 of 44 tasks
(90.91\%) at Step~1. Iterative target reselection improves
accuracy to 95.45\%, 97.73\%, and 100\% over the next three steps.
These results show that dynamic reflection reduces one-shot node selection errors. Instead of relying on a fixed initial prediction, \sys can revise the target node based on the updated UTG and execution feedback, making target-node errors recoverable during planning.

\section{Discussion}
\label{sec:discussion}

\noindent\textbf{Symbolic Planning as a Remedy for LLM Hallucination.}
Our study demonstrates that integrating classical symbolic planning with LLM-based agents substantially mitigates long-horizon planning failures. Unlike end-to-end LLM reasoning, which often suffers from hallucination and myopic exploration, symbolic planners provide \textit{global guarantees} on correctness and optimality. This synergy leverages the complementary strengths of both paradigms: the interpretability and reliability of symbolic reasoning, and the perception and linguistic versatility of LLMs.

\noindent\textbf{Generalization.} \sys's methodology extends beyond Android UI automation to any domain modeled by state--transition graphs, including web~\cite{zhou2023webarena} and robotics~\cite{zhu2025psalm}. By substituting the UTG with analogous structures (e.g., DOM), \sys offers a blueprint for a unified symbolic--neural framework for general UI reasoning.

\noindent\textbf{Multi-Goal Automation.}
 Our current implementation simplifies UI automation tasks into single-goal navigation problems, where the objective is to reach a single target UI. However, many practical user tasks are inherently multifaceted and require achieving a sequence of sub-goals. For instance, a task like \textit{``Add an item to the shopping cart and then proceed to checkout''} involves successfully reaching the item's page (goal 1) and subsequently navigating to the checkout screen (goal 2).

Extending \sys to handle multi-goal scenarios would involve evolving the Node Selector into a more sophisticated \textit{``Goal Decomposer''} capable of parsing a complex natural language instruction into an ordered list of target UI nodes $\{u_{target\_1}, u_{target\_2}, \dots, u_{target\_n}\}$. Subsequently, the Plan Generator would leverage the native ability of PDDL to support multiple goal predicates, enabling the generation of a single, cohesive plan that traverses the UTG to satisfy all sub-goals. This effectively extends \sys to a hierarchical planning paradigm. However, this approach also raises new research questions regarding goal ordering and dependency resolution, especially when goals become infeasible at runtime. 

\section{Related Work}

\subsection{UI Automation}  %
Traditionally, research in UI automation has centered on \textit{automated testing}, where the primary objective is to systematically explore an app's UIs to discover bugs~\cite{ran2024guardian,hu2011automating,lai2019goal}, or security vulnerabilities~\cite{shahriar2009automatic,ye2009intelligent,moura2023cytestion,liu2020maddroid}. 
Graph learning methods~\cite{ju2023graphpatcher, ju2022adaptive, qian2022co, ju2022grape, zhao2021multi} have provided powerful capabilities for modeling structured entities and their complex dependencies, offering a natural foundation for representing interconnected UI elements and interactions.
With more recent advances of LLMs, the focus has expanded to \textit{task-driven GUI agents}, which aim to complete specific real-world tasks rather than maximizing the exploration coverage~\cite{wen2024autodroid,rawles2024androidworld,chen2024empathy}.

\subsection{LLM + Symbolic Planner}
Prior work demonstrates the mutual benefits of combining LLMs~\cite{ye2025llms4all} with symbolic planners. One prominent direction involves using LLMs to formalize natural language problems into PDDL, allowing classical solvers to guarantee reliable execution~\cite{liu2023llm+, dagan2023dynamic, guan2023leveraging, valmeekam2023planning}. Conversely, other approaches use LLMs to augment symbolic planners by interpreting environment feedback~\cite{zhu2025language,zhu2025psalm}. \sys synthesizes insights from both directions: it employs a symbolic planner to guide the LLM's decision-making, while simultaneously leveraging the LLM to interpret UI feedback and refine the symbolic planning space.

\subsection{UI Transition Modeling}
Existing works construct UTGs (or similar concepts) using various methods across multiple platforms. One primary approach is \textit{dynamic analysis}, where the application is executed to observe real transitions. This is seen in early, multi-platform (e.g., web, mobile) systems like GUITAR~\cite{nguyen2014guitar} and more recent tools such as GUI-Xplore~\cite{sun2025gui} and Autodroid~\cite{wen2024autodroid}. The other approach is \textit{static analysis}, which infers control flow from the app's code, such as $A^3E$'s activity transition graph~\cite{dfsandroid} and Gator's window transition graph~\cite{yang2018static}.  \sys constructs the UTG to model UI transitions by synergistically combining both static and dynamic techniques.

\section{Conclusion}
\label{sec:conclusion}

We introduced \sys, an agentic framework for long-horizon planning in UI automation. By modeling an app's transition structure as a UTG and leveraging an external symbolic planner, \sys provides LLM-based agents with a globally optimal, high-level plan, effectively mitigating hallucination that causes common automation failure. Our evaluation on the AndroidWorld benchmark demonstrates that \sys can be integrated as a plug-and-play module to substantially improve the success rates of state-of-the-art UI agents. We believe \sys lays the foundation for future research into neuro-symbolic planning paradigms for creating more robust and reliable UI agents for complex UI automation tasks.

\section*{Limitations}

While \sys effectively bridges global planning with local reasoning, several challenges remain. 
First, constructing accurate UTGs still depends on program analysis quality. Static analysis may include infeasible edges, while dynamic exploration may yield incomplete coverage. 
Second, deploying \sys on a new app requires obtaining, decompiling, and analyzing its APK when source code is unavailable. Although this preprocessing can be performed offline, it limits immediate deployment on previously unanalyzed apps.  
Third, translating between natural language and PDDL representations introduces an additional layer of abstraction; errors in this translation can propagate through the pipeline. 
Finally, classical planners assume deterministic transitions, yet real-world GUIs often contain stochastic behaviors (e.g., pop-ups or async UI updates), which may lead to plan divergence during execution. 
Addressing these issues requires tighter integration between the planner and the environment to support real-time re-planning and uncertainty handling.

\section*{Ethical Considerations}
Automated UI agents that can execute complex tasks on real applications must be deployed with care. 
Potential misuse, such as automating sensitive or privacy-related operations without explicit consent, highlights the need for transparent auditing and permission control. 
In our implementation, all experiments are conducted on benign benchmark apps under controlled environments.

\section*{Acknowledgements}
This work was partially supported by the NSF under grants IIS-2533550, IIS-2321504, IIS-2217239, CNS-2426514, and CMMI-2146076, Notre Dame Strategic Framework Research Grant (2025), and Notre Dame Poverty Research Package (2025). Any expressed opinions, findings, and conclusions or recommendations are those of the authors and do not necessarily reflect the views of the sponsors.

\bibliography{emnlp2023}

\clearpage
\appendix
\label{sec:appendix}

\section{Implementation} 
At the time of our experiments, MobileUse and DroidRun claimed $0.91$ on the official AndroidWorld leaderboard. 
However, the exact configurations used to obtain these results (e.g., backbone LLMs, enabling reasoning or vision capabilities) were not publicly disclosed. 
To ensure a fair comparison, we rerun each agent using three state-of-the-art LLMs: GPT-5, Gemini-3, and Grok 4, under multiple parameter configurations.  
Runs that terminated due to environment or infrastructure errors before the agent could act (e.g., emulator crashes, API timeouts) are excluded from the metrics. This affected 6 tasks for T3A. Success rate is computed over the remaining tasks.
We adopt the configuration with the best performance for our evaluation.

The static UTG is constructed using ICCBot~\cite{yan2022iccbot} and FlowDroid~\cite{arzt2014flowdroid}. 
For the classical planner, we employ Fast Downward with an $A^*$ search algorithm to find optimal paths. 
 To ensure the reliability of our findings, all experiments were conducted three times, and we report the average values.

 \section{Case Study}
  \label{sec:case-study}

\begin{figure*}[tp]
      \centering

      \textbf{Observed UI States}\\[5pt]

      \begin{minipage}[t]{0.30\textwidth}
          \centering
          \includegraphics[width=0.78\linewidth]{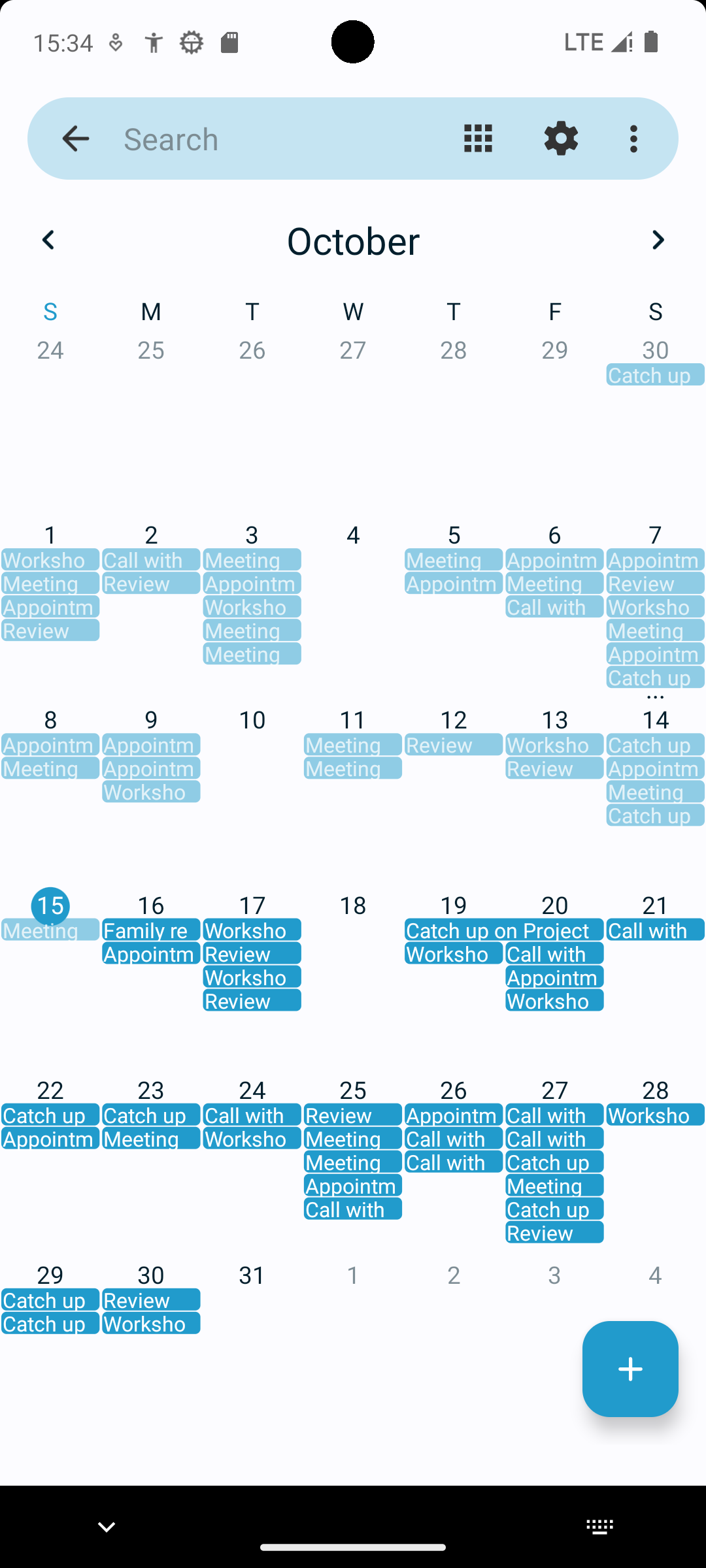}\\[2pt]
          \small (a) Initial UI
      \end{minipage}
      \hfill
      \begin{minipage}[t]{0.30\textwidth}
          \centering
          \includegraphics[width=0.78\linewidth]{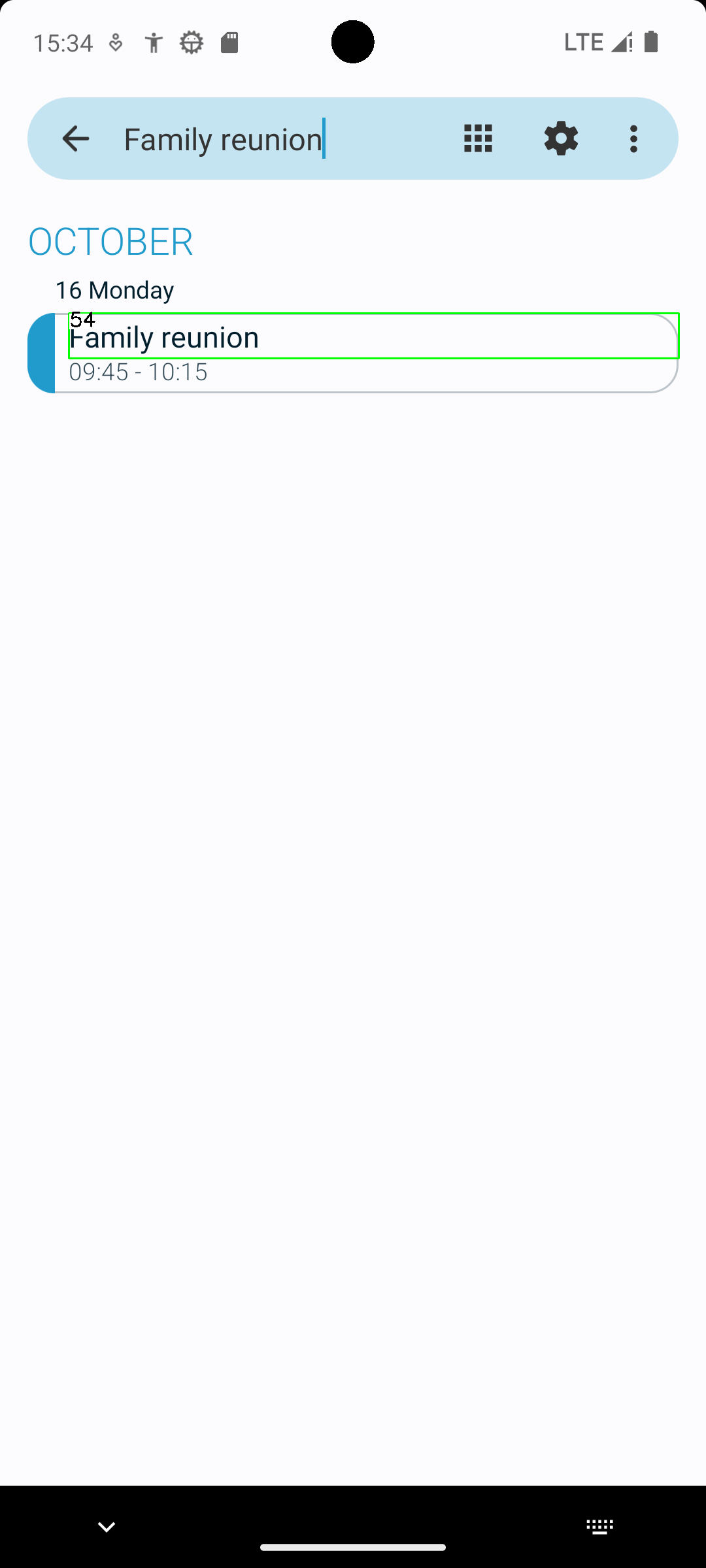}\\[2pt]
          \small (b) Intermediate UI
      \end{minipage}
      \hfill
      \begin{minipage}[t]{0.30\textwidth}
          \centering
          \includegraphics[width=0.78\linewidth]{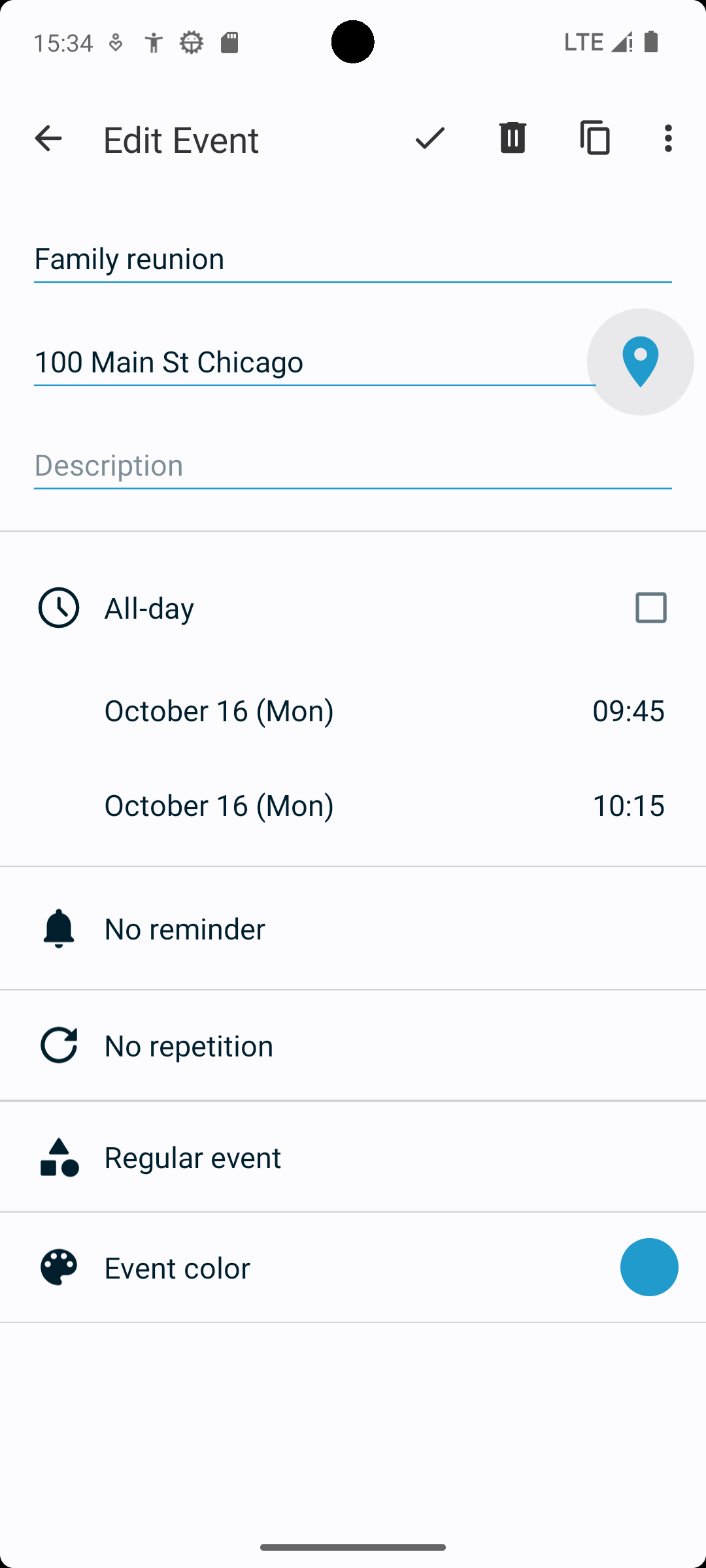}\\[2pt]
          \small (c) Target UI
      \end{minipage}

      \vspace{5pt}
      \(\Downarrow\)
      \vspace{5pt}

  \fcolorbox{blue!55!black}{blue!5}{%
      \parbox{0.94\textwidth}{%
          \centering
          \textbf{UTG-Based Plan}\\[6pt]

          \begin{minipage}[c]{0.39\linewidth}
              \small
              \textbf{Current node:}
              \texttt{MainActivity}\\
              \textbf{Target node:}
              \texttt{EventActivity}\\
              \textbf{Trigger widget:}
              Event item ``Family reunion''\\
              \textbf{Confidence:} 90\%
          \end{minipage}
          \hfill
          \begin{minipage}[c]{0.54\linewidth}
              \centering
              \small
              \textbf{Optimal transition}\\[4pt]
\texttt{MainActivity} to \texttt{EventActivity}\\
on ``Click'' the ``\texttt{EventItem}'' widget with content description ``Family reunion.''\\[5pt]
          \end{minipage}
      }%
  }

      \vspace{5pt}
      \(\Downarrow\)
      \vspace{5pt}

      \fcolorbox{green!45!black}{green!5}{%
          \parbox{0.94\textwidth}{%
              \centering
              \textbf{Actions Executed by the UI Agent}\\[7pt]

              \begin{minipage}[c]{0.28\linewidth}
                  \centering\small
                  \texttt{input\_text(}\\
                  \texttt{"Family reunion",}\\
                  \texttt{search\_field)}
              \end{minipage}
              \(\longrightarrow\)
              \begin{minipage}[c]{0.28\linewidth}
                  \centering\small
                  \texttt{click(}\\
                  \texttt{event\_item=}\\
                  \texttt{"Family reunion")}
              \end{minipage}
              \(\longrightarrow\)
              \begin{minipage}[c]{0.28\linewidth}
                  \centering\small
                  \texttt{answer(}\\
                  \texttt{location=}\\
                  \texttt{"100 Main St Chicago")}
              \end{minipage}
          }%
      }
      \caption{Step-by-step case study.}
      \label{fig:calendar-case-study}
  \end{figure*}

  Figure~\ref{fig:calendar-case-study} illustrates a successful execution for the task: \emph{``What is the location of my Family reunion event?''} The agent begins on the calendar screen and
  searches for the event by name. Once the matching event appears, it selects the result and reaches the event-details screen, where the location, ``100 Main St Chicago,'' is visible.

  The simplified plan provides the high-level transition from the main calendar to the event-details screen. The UI agent realizes this plan through concrete actions: entering the event name,
  selecting the matching result, and returning the displayed location. This example shows how the plan guides navigation while the agent handles the specific interactions required by the current UI.

\section{Details on Static UTG Construction}

We construct the UTG $G = (\mathcal{U}, \mathcal{T}, \epsilon)$ via a three-step static analysis of the source code of an Android app.

\noindent\textbf{UI Identification ($\mathcal{U}$).}
First, we identify the set of all possible UIs, $\mathcal{U}$. In the context of Android applications, a UI state generally corresponds to one of three primary components:
\begin{itemize}[noitemsep, topsep=1pt, partopsep=1pt, leftmargin=*]
    \item \textbf{Activity:} An \incode{Activity} represents a single, focused screen with a user interface. It serves as the entry point for interacting with the user and usually occupies the entire display window.
    \item \textbf{Dialog:} A \incode{Dialog} is a small window that appears in front of the current Activity. It captures user focus for critical decisions or additional input without navigating away from the underlying screen.
    \item \textbf{Fragment:} A \incode{Fragment} represents a reusable portion of the user interface within an Activity. Fragments have their own lifecycle and input handling, allowing for dynamic and flexible UI designs (e.g., tabbed views or sidebars).
\end{itemize}

\begin{lstlisting}[language=Java, caption={Code snippet of an activity.}, label={lst:node_extraction}]
public class CalendarActivity extends Activity {
    @Override
    protected void onCreate(Bundle savedInstanceState) {
        super.onCreate(savedInstanceState);
        // We identify this onCreate call as the creation of the activity.
        setContentView(R.layout.activity_calendar);
    }
}
\end{lstlisting}

\noindent\textbf{Transition Extraction ($\mathcal{T}$).}
Next, we construct the set of directed edges $\mathcal{T}$. We analyze methods containing navigation-related API calls to determine target UIs.
\begin{itemize}[noitemsep, topsep=1pt, partopsep=1pt, leftmargin=*]
    \item \textbf{Intents:} For \incode{startActivity(Intent)}, we resolve the target component class from the \incode{Intent} constructor or \incode{setClass} method.
    \item \textbf{Fragment Transactions:} For internal transitions, we track the target fragment class used in \incode{FragmentTransaction}.
\end{itemize}
If a source UI $u_i$ (e.g., \incode{CalendarActivity}) contains a reachable call that launches a target UI $u_j$ (e.g., \incode{EventActivity}), we add an edge $(u_i, u_j) \in \mathcal{T}$.

\noindent\textbf{Edge Labeling ($\epsilon$).}
Finally, we define the edge-labeling function $\epsilon$ by linking each transition $(u_i, u_j)$ back to the specific widget interaction that triggers it. We perform a reverse reachability analysis from the navigation call (identified in Step 2) to the enclosing event listener (e.g., \incode{onClick}).

This allows us to characterize the action $a = (w, e)$ where $w$ is the widget bound to the listener and $e$ is the event type. As shown in Listing~\ref{lst:edge_labeling}, the button \incode{btnAddEvent} is identified as the widget $w$ responsible for the transition to \incode{EventActivity}.

\begin{lstlisting}[language=Java, caption={Linking a transition to an action $a=(w, \mathtt{click})$.}, label={lst:edge_labeling}]
Button btnAddEvent = findViewById(R.id.btn);
// The event listener defines the action type click
btnAddEvent.setOnClickListener(new View.OnClickListener() {
    @Override
    public void onClick(View v) {
        // We link this navigation to widget btnAddEvent
        // Creates UTG edge 
        Intent intent = new Intent(CalendarActivity.this, EventActivity.class);
        startActivity(intent);
    }
});
\end{lstlisting}

\section{Additional Figures and Tables}
\label{app:figtable}

\begin{table}[htbp]
\centering
\small
\begin{tabular}{@{}llll@{}}
\toprule
\multicolumn{1}{c}{\textbf{App}} & \multicolumn{1}{c}{\textbf{Tasks}} & \multicolumn{1}{c}{\textbf{Nodes}} & \multicolumn{1}{c}{\textbf{Edges}} \\ \midrule
VLC                              & 3                                  & 85                                 & 190                                \\
Simple Calendar Pro                        & 17                                 & 24                                 & 25                                 \\
Tasks                            & 5                                  & 86                                 & 127                                \\
Markor                           & 16                                 & 14                                 & 17                                 \\
OsmAnd                           & 3                                  & 152                                & 508                                \\ \bottomrule
\end{tabular}
\caption{Statistics of the AndroidWorld subset used in evaluation.}
\label{tab:evaldataset}
\end{table}

\begin{table}[htbp]
\centering
\small %
\setlength{\tabcolsep}{4pt} %
\begin{tabular}{@{}lcc@{}}
\toprule
\textbf{Method} & \textbf{SR (\%)} & \textbf{Improvement} \\
\midrule

DroidRun & 77.59 & -- \\
\rowcolor{gray!15}
\hspace{1em} + \sys & \textbf{78.45} & \textbf{+0.86} \\
\addlinespace

MobileUse & 51.72 & -- \\
\rowcolor{gray!15}
\hspace{1em} + \sys & \textbf{62.93} & \textbf{+11.21} \\
\addlinespace

T3A & 62.07 & -- \\
\rowcolor{gray!15}
\hspace{1em} + \sys & \textbf{63.79} & \textbf{+1.72} \\
\addlinespace

OpenAI CUA & 80.17 & -- \\
\rowcolor{gray!15}
\hspace{1em} + \sys & \textbf{82.76} & \textbf{+2.59} \\

\bottomrule
\end{tabular}
\caption{Performance of \sys and baselines on the full AndroidWorld benchmark.}
\label{tab:main_results_fullbenchmark}
\end{table}

\begin{figure}[htbp]
\centering
\begin{lstlisting}[language=lisp, caption={Example problem PDDL for the Simple Calendar Pro app with the task \textit{Change the time zone}.}, label={lst:problem_pddl}]
(define (problem change-time-zone)
  (:domain utg-automation)

  (:objects
    SplashActivity - node
    MainActivity - node
    EventActivity - node
    SettingsActivity - node
    AboutActivity - node
    TaskActivity - node
    SelectTimeZoneActivity - node
    ManageEventTypesActivity - node
    WidgetListConfigureActivity - node
    ContributorsActivity - node
    FAQActivity - node
    LicenseActivity - node
  )

  (:init
    (at SplashActivity)

    (goal-node SelectTimeZoneActivity)

    (connected SplashActivity MainActivity)
    (connected MainActivity EventActivity)
    (connected MainActivity SettingsActivity)
    (connected MainActivity AboutActivity)
    (connected MainActivity TaskActivity)
    (connected EventActivity EventActivity)
    (connected EventActivity SelectTimeZoneActivity)
    (connected SettingsActivity ManageEventTypesActivity)
    (connected SettingsActivity WidgetListConfigureActivity)
    (connected AboutActivity ContributorsActivity)
    (connected AboutActivity FAQActivity)
    (connected AboutActivity LicenseActivity)
    (connected TaskActivity TaskActivity)
  )

  (:goal
    (goal-achieved SelectTimeZoneActivity)
  )
)

\end{lstlisting}
\end{figure}

\begin{figure}[htbp]
\label{code:NodeSelectorPrompt}
\centering
\begin{lstlisting}[language=lisp, caption={Prompt template of Node Selector.}, label={lst:prompt_nodeSelector}]
Given the user goal: "{user_goal}".

Your task is to select the most relevant UTG node that can help achieve this goal.

Previous selected target node:
{previous_target_node}

Available UTG nodes:
{chr(10).join(nodes_information)}

Selection rules:
1. If the previous target node is still suitable for the user goal, keep it.
2. If the previous target node is incorrect, unreachable, or less relevant than another node, revise it.
3. Select the node based on both the user goal and the current UTG information.
4. Prefer a concrete UI state that directly supports task completion, rather than an intermediate navigation screen.

Return your response in JSON format:
{{
    "node": "selected_node_id",
    "decision": "keep_or_revise",
    "confidence": 0.8,
    "reasoning": "brief explanation"
}}

Return ONLY the JSON, no other text.
\end{lstlisting}
\end{figure}

\begin{figure}[htbp]
\centering
\begin{lstlisting}[language=lisp, caption={The plan generated by Fast Downward $A^*$ for the Simple Calendar Pro app with the task \textit{Add a new event type named ``1-on-1 meeting''}.}, label={lst:plan_pddl}]
(navigate  SplashActivity MainActivity)
(navigate MainActivity SettingsActivity)
(navigate SettingsActivity ManageEventTypesActivity)
; cost = 3 (unit cost)
\end{lstlisting}
\end{figure}

\begin{figure*}[tp]
    \centering
    \includegraphics[width=\textwidth]{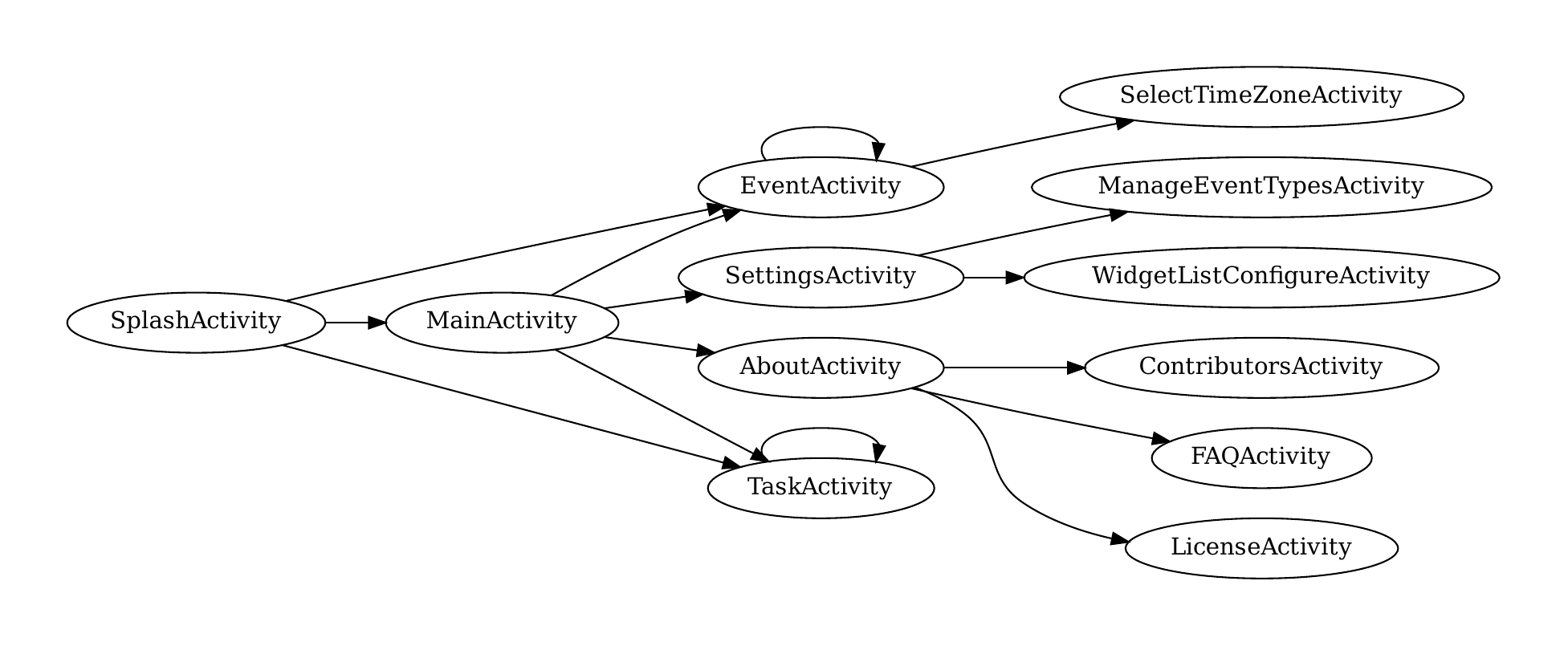}
    \caption{Graphviz visualization of the UTG of Simple Calendar Pro in AndroidWorld. An activity is a unit of Android UI~\cite{activity}. Edge labels are removed for visual clarity. }
    \label{fig:utgCalendar}
\end{figure*}

\begin{table*}[tp]
\centering
\resizebox{\linewidth}{!}{%
\begin{tabular}{|l|l|}
\hline
\multicolumn{1}{|c|}{\textbf{Category}}                             & \multicolumn{1}{c|}{\textbf{Content}}                                                                                                                                                                                                                                                                                                                                                                                                                                                                                                                                                                                                                                                                                                                                   \\ \hline
\begin{tabular}[c]{@{}l@{}}Natural \\ Language \\ Instructions\end{tabular} & \begin{tabular}[c]{@{}l@{}}--- UTG Navigation Guide ---\\ Current UI: MainActivity\\ \\ --- NAVIGATION PLAN FOR YOUR GOAL ---\\ Goal Analysis: Based on your goal, the system identified these target destinations:\\   \textbullet{} ManageEventTypesActivity\\ Confidence: 95\%\\ \\ OPTIMAL PATH (Follow these steps in order):\\ Step 1: Navigate from MainActivity to SettingsActivity\\ Step 2: Navigate from SettingsActivity to ManageEventTypesActivity\\ IMMEDIATE NEXT ACTION:\\   $\rightarrow$  on click the ImageView widget with content-description ``more options''  via API call \\ ``virtualinvoke r0.<android.content.Context: void startActivity(android.content.Intent)>(r1) ()''\\   This will take you to: SettingsActivity\\ Total steps in optimal path: 2\end{tabular} \\ \hline
\multicolumn{1}{|c|}{Usage Tips}                                    & \begin{tabular}[c]{@{}l@{}}This path was computed using PDDL planning for guaranteed optimality\\ --- USAGE TIPS ---\\ \textbullet{} If a NAVIGATION PLAN is shown above, follow it step by step for optimal path\\ \textbullet{} The plan was computed using formal PDDL planning algorithms\\ \textbullet{} If no plan exists, the target is unreachable from current location\\ \textbullet{} Some UI elements may not be visible - scroll if needed\end{tabular}                                                                                                                                                                                                                                                                                                                                                    \\ \hline
Fallback                                                            & \begin{tabular}[c]{@{}l@{}}--- ALL AVAILABLE NAVIGATION OPTIONS FROM HERE ---\\ \textbullet{} TO REACH: EventActivity\\   $\rightarrow$  on click ImageButton widget with content-description ``New Event'' via  API call \\ ``virtualinvoke r0.<android.content.Context: void startActivity(android.content.Intent)>(r1) ()"\\ \textbullet{} TO REACH: SettingsActivity\\   $\rightarrow$  on click Button widget with content-description ``Settings'' via API call \\ ``virtualinvoke r0.<android.content.Context: void startActivity(android.content.Intent)>(r1) ()"\\ \textbullet{} TO REACH: TaskActivity\\   $\rightarrow$  on click ImageButton widget via api call ``virtualinvoke \\ r2.<android.content.Context: void startActivity(android.content.Intent)>(\\ r3) ()''\\ --- End Navigation Guide ---\end{tabular}                  \\ \hline
\end{tabular}
}
\caption{Example natural language prompt for the Simple Calendar Pro app with the task \textit{Add a new event type named `1-on-1 meeting'}.}
\label{tab:utgGuidePrompt}
\end{table*}

\end{document}